\documentclass[conference]{IEEEtran}

\newif\ifArXivVersion
\ArXivVersiontrue

\usepackage[left=0.625in,right=0.625in,top=1.8cm,bottom=1.05in]{geometry}

\newif\ifreview
\reviewfalse

\newif\iflargeauthor
\largeauthortrue

\usepackage{cite}
\usepackage{amsmath,amssymb,amsfonts}
\usepackage{algorithmic}
\usepackage{graphicx}
\usepackage{tikz}
\usepackage{siunitx}  %
\usepackage{hyperref}
\usepackage{float,booktabs,xcolor,textcomp}

\def\BibTeX{{\rm B\kern-.05em{\sc i\kern-.025em b}\kern-.08em
    T\kern-.1667em\lower.7ex\hbox{E}\kern-.125emX}}

\newcommand{\numzero}[1]{\num[minimum-integer-digits=2]{#1}}

\begin{document}

\title{Demystifying Trajectory Recovery From Ash:\\An Open-Source Evaluation and Enhancement
}
\iflargeauthor
\author{
\IEEEauthorblockN{Nicholas D'Silva}
\IEEEauthorblockA{\textit{University of New South Wales} \\
Sydney, Australia \\
n.dsilva@unsw.edu.au}
\and
\IEEEauthorblockN{Toran Shahi}
\IEEEauthorblockA{\textit{University of New South Wales} \\
Sydney, Australia \\
toranjungshahi@gmail.com}
\and
\IEEEauthorblockN{Øyvind Timian Dokk Husveg}
\IEEEauthorblockA{\textit{University of New South Wales} \\
Sydney, Australia \\
timian@husveg.net}
\and
\IEEEauthorblockN{Adith Sanjeeve}
\IEEEauthorblockA{\textit{University of New South Wales} \\
Sydney, Australia \\
adithsanjeeve1331@gmail.com}
\and
\IEEEauthorblockN{Erik Buchholz}
\IEEEauthorblockA{\textit{University of New South Wales} \\
\textit{CSIRO's Data61, Cyber Security CRC}\\
e.buchholz@unsw.edu.au
}
\and
\IEEEauthorblockN{Salil S. Kanhere}
\IEEEauthorblockA{\textit{University of New South Wales} \\
Sydney, Australia \\
salil.kanhere@unsw.edu.au}
}

\fi

\ifArXivVersion
    \newcommand\copyrighttext{%
      \footnotesize \textcopyright 2024 IEEE. To appear in the proceedings of the 17th International Conference on Security of Information and Networks (SIN'24).
      This is the author’s version of the work. It is posted here for your personal use. Not
for redistribution. 
      }
    
    \newcommand\copyrightnotice{%
    \begin{tikzpicture}[remember picture,overlay]
    \node[anchor=south,yshift=20pt] at (current page.south) {\fbox{\parbox{\dimexpr0.75\textwidth-\fboxsep-\fboxrule\relax}{\copyrighttext}}};
    \end{tikzpicture}%
    }
\fi

\maketitle

\begin{abstract}
Once analysed, location trajectories can provide valuable insights beneficial to various applications, including urban planning, market analysis, and public health surveillance. However, such data is also highly sensitive, rendering them susceptible to privacy risks in the event of mismanagement, for example, revealing an individual's identity, home address, or political affiliations. Hence, ensuring that privacy is preserved for this data is a priority. One commonly taken measure to mitigate this concern is aggregation. Previous work by Xu et al. in [Trajectory Recovery From Ash: User Privacy Is NOT Preserved in Aggregated Mobility Data (2017)] shows that trajectories are still recoverable from anonymised and aggregated datasets. However, the study lacks implementation details, obfuscating the mechanisms of the attack. Additionally, the attack was evaluated on commercial non-public datasets, rendering the results and subsequent claims unverifiable. This study reimplements the trajectory recovery attack from scratch and evaluates it on two open-source datasets, detailing the preprocessing steps and implementation. Results confirm that privacy leakage still exists despite common anonymisation and aggregation methods but also indicate that the initial accuracy claims may have been overly ambitious. We release all code as open-source to ensure the results are entirely reproducible and, therefore, verifiable. Moreover, we propose a stronger attack by designing a series of enhancements to the baseline attack. These enhancements yield higher accuracies by up to 16\%, providing an improved benchmark for future research in trajectory recovery methods. Our improvements also enable online execution of the attack, allowing partial attacks on larger datasets previously considered unprocessable, thereby furthering the extent of privacy leakage. The findings emphasise the importance of using strong privacy-preserving mechanisms when releasing aggregated mobility data and not solely relying on aggregation as a means of anonymisation.
\end{abstract}


\ifArXivVersion
    \copyrightnotice
\fi

\section{Introduction}\label{intro}
The domain of human mobility data collection and analysis is of growing importance in both academic and industrial spheres. Some contexts where such data proves beneficial include urban planning, pandemic response analysis, and marketing. Advancements in technology, particularly with personal mobile devices, have facilitated the collection of human mobility data in increasingly higher quantities and quality. However, collecting this data also comes with significant privacy concerns due to the risk of inference of sensitive information. For instance, exploits or mismanagement can lead to identity theft or harassment due to publicised home addresses or personal beliefs. One commonly taken measure to mitigate this concern after anonymisation is aggregation. For example, this may involve transforming a dataset of individual trajectories into a dataset showing the number of individuals within a set of predefined locations over some period of time.

A critical examination of aggregated mobility datasets reveals a serious vulnerability: the ease of re-identification. This concern is highlighted in the work of Xu et al.~\cite{Xu_2017}, who demonstrate that aggregated mobility data, despite statistically obfuscating individual records, can be de-anonymised by reconstructing trajectories and potentially revealing sensitive individual information. The authors describe their design as an ``elementary but effective attack system to reveal the privacy leakage in aggregated mobility datasets''~\cite{Xu_2017}. They evaluate their attack on two real-world but inaccessible datasets and report accuracies of up to \SI{91}{\%}.

Because of their highly sensitive nature, datasets such as the commercial ones used in the work by Xu et al. are not publicly available, limiting further research possibilities. In conjunction with the fact that the study does not provide any implementation details about their attack, it makes their results irreproducible and renders the claims unverifiable. Intending to increase clarity and transparency in this area, we reimplement the attack they present from scratch, design and implement further enhancements to the attack, and perform evaluations on two public open-source datasets, namely GeoLife~\cite{GeoLife} and Porto Taxi~\cite{Porto_Taxi}, and release all our code as open-source\footnote{
\ifreview
\label{code}\url{https://github.com/ANONYMIZED_FOR_REVIEW}
\else
\label{code}\url{https://github.com/ndsi6382/Trajectory_Recovery}
\fi
\label{repo}}.
Using more accessible datasets allows us to explain with greater transparency the specific characteristics inherent to the aggregated input datasets supposed to contain privacy leakages. We detail our preprocessing methodology for each of our chosen datasets to ensure our results are reproducible and verifiable. We are convinced that our reimplementation and explanations further clarify the attack process, making it more accessible for further research\footnote{
An alternative implementation~\cite{Tu_2018} of a paper based on~\cite{Xu_2017} exists. However, our version includes the processed open datasets, an abstracted evaluation module, a detailed walk-through of the attack process, and our proposed enhancements, features absent in that existing implementation.
}.
Additionally, the enhancements we propose provide a more accurate baseline against which future researchers can benchmark their work, particularly within the field of deep learning. The enhancements also permit the attack to be run online, significantly increasing its accessibility. Partial attacks can be conducted on larger datasets that were previously considered unprocessable by the baseline attack, furthering the extent of the privacy leakage.

\noindent\textbf{Contributions.} This work makes the following contributions to the field of location trajectory privacy:
\begin{enumerate}
    \item Evaluated the validity of the results and claims in~\cite{Xu_2017}:
    \renewcommand{\labelenumii}{\alph{enumii})}
    \begin{enumerate}
        \item Reimplemented algorithms from \cite{Xu_2017}.
        \item Preprocessed two publicly available open-source datasets and applied the algorithm to each.
    \end{enumerate}
    \item Designed, implemented, and evaluated a series of enhancements to the baseline algorithm:
    \renewcommand{\labelenumii}{\alph{enumii})}
    \begin{enumerate}
        \item Developed a stronger attack against which future research can use as a baseline.
        \item Showed that our described online methodology allows adversaries to attack larger datasets, furthering the privacy leakage.
    \end{enumerate}
    \item Encouraged further research, with an emphasis on clarity and transparency:
    \renewcommand{\labelenumii}{\alph{enumii})}
    \begin{enumerate}
        \item Released all source code, data, results, and supplementary resources as open-source\footref{repo}, thus ensuring our results are reproducible and verifiable.
        \item Produced guides detailing our preprocessing and algorithm implementations.
        \item Packaged all algorithms as a Python module with full documentation.
    \end{enumerate}
\end{enumerate}

\noindent\textbf{Organisation.} In Section~\ref{preliminaries}, we contextualise our work, outline a threat model, formally define the problem, summarise the attack from~\cite{Xu_2017}, and consequently make clarifying statements about the required properties of the aggregated dataset for the attack to function as intended. We describe our enhancements in Section~\ref{enhancements}. In Section~\ref{data}, we analyse each open-source dataset and explain our preprocessing methodology. Implementation details are provided in Section~\ref{implementation}. We evaluate and discuss results in Section~\ref{evaluation}, mention future directions in Section~\ref{future}, and provide concluding remarks in Section~\ref{conclusion}.

\section{Preliminaries}\label{preliminaries}
\subsection{Related Work}\label{background}
As the domain of mobility data analysis has grown, so has the emphasis on protecting the privacy of individuals, giving rise to privacy mechanisms~\cite{Primault_2019, Miranda-Pascual2023, Buchholz_2024} based on $k$-anonymity~\cite{Sweeney_2002} and differential privacy (DP)~\cite{Dwork_2006}. In recent decades, studies in human mobility have revealed that humans exhibit exceptionally distinctive patterns~\cite{Gonzalez_2008, Sui_2016, Montjoye_2013}. Their trajectories, despite being anonymised, are largely unique and thus pose a risk for re-identification. Consequently, several attacks have been designed targeting anonymised location data, exposing that re-identification is possible with external cross-referenced information~\cite{Shokri_2011}, and even without~\cite{Montjoye_2013, Gambs_2014}.

To alleviate concerns of privacy leakage, data collectors and providers often aggregate sensitive information, including locations, before publication. Many recent attacks targeting aggregated location data are membership inference attacks that identify whether an individual's data is included in the supposedly anonymised dataset, giving adversaries access to sensitive information. Examples include the Knock-Knock attack by Pyrgelis et al.~\cite{Pyrgelis_2017}, with more recent developments by Zhang et al.~\cite{Zhang_2020}, and Guan and Guépin et al.~\cite{Guan_2024} showing that less or zero prior knowledge is required from adversaries. Other recent developments include individual reconstruction attacks~\cite{Shao2020, RAoPT} that target reconstructing the original trajectories from a protected (for example, with DP) trajectory dataset. Contrary to these works, the attack evaluated and improved upon in this study reconstructs trajectories from a dataset aggregated by location rather than a protected trajectory dataset. The potential consequences of this are detailed below.

\subsection{Threat Model}\label{threat_model}
The \textit{data owner} is an entity that collects data from \textit{individuals}, such as a mobile phone network provider gathering connection information. 
The data owner plans to share an aggregated dataset for the \textit{data recipient}'s use.
As per the baseline work~\cite{Xu_2017}, we consider the data owner trustworthy. Individuals trust the data owner to adequately anonymise their data (through aggregation).
However, neither the data owner nor the individuals trust the data recipient, who acts as an \textit{honest-but-curious adversary} in this threat model. 
The adversary aims to extract as much information as possible about the individuals in the aggregated dataset. 

The considered attack~\cite{Xu_2017} allows the adversary to recover the trajectories of contained individuals without requiring any background knowledge. 
While this set of trajectories is still de-identified, existing effective re-identification techniques~\cite{Montjoye_2013} or trajectory user linking~\cite{Gao_2017,marc2020} can further be applied, exposing privacy leakage. Note the three-stage design of this attack (see Section~\ref{attack_summary}) requires the entire dataset to be processed before re-identification can occur. Therefore, the baseline attack applies only to datasets of a size that can be processed within a realistic timeframe.

In contrast, adversaries can conduct our enhanced attack (detailed in Section~\ref{enhancements}) online, i.e. it processes data and outputs results sequentially without requiring the entire input upfront~\cite{Karp_1992}. Given the same computational resources, this allows the adversary to target subsets of aggregated mobility datasets that were previously considered too large for the baseline attack to process. Noting that re-identification risk only slowly decreases proportionally to the number of individuals in the dataset~\cite{Farzanehfar_2021}, this method increases the privacy leakage. The modified order of computation allows for data to be processed in chronological order and batches no smaller than one day. Noting that parallel algorithms exist for solving the Linear Sum Assignment problem~\cite{Date_2016}, the remaining computational bottleneck relates to the number of time steps covered by the dataset (see Sections~\ref{attack_summary} and~\ref{enhancements}). With the online method, intermittent results can be retrieved either during execution, or after an early (manual) termination of the algorithm. This allows for contiguous sub-trajectories to be used for re-identification instead of requiring all time steps to be processed first, thus removing the dataset size limitation of the baseline work mentioned above. This yields an increased attack surface.

\subsection{Definitions}\label{definitions}
\noindent\textbf{Dataset.} The baseline attack presented in~\cite{Xu_2017} is formulated as a deterministic algorithm for which we define the problem as follows. An anonymised, aggregated dataset $D \in \mathbb{N}^{t \times m}$ contains $t$ records, where each record $r_i \in \mathbb{N}^{1 \times m}$ for $1 \le i \le t$ contains the number of individuals in each of the $m$ locations at the $i$th time step. Given $D$, output a set $S \in \mathbb{R}^{n \times t \times 2}$ of reconstructed trajectories for each of the $n$ individuals captured in $D$, where each trajectory $v_j \in \mathbb{R}^{1 \times t \times 2}$ for $1 \le j \le n$ contains the two-dimensional location coordinates for the $j$th individual, for every time step in chronological order.

\noindent \textbf{Hungarian algorithm.} The attack makes extensive use of the Hungarian algorithm (also known as the ``Munkres'' or ``Kuhne-Munkres'' algorithm) to solve the square assignment problem~\cite{Kuhn_1955}. Given a square matrix $C \in \mathbb{R}^{n \times n}$, where rows represent assignees and columns represent assignments, each element $c_{i,j} \in C$ is defined as the cost of assigning $i$ to $j$. The objective is to determine a one-to-one matching of assignees to assignments that results in the optimal (minimum or maximum) total cost. This is often alternatively described as the Linear Sum Assignment problem~\cite{Dantzig_1965}, where $n$ elements must be selected from the square matrix, subject to the constraints that exactly one element is selected from each row and each column, and to optimise the total sum. The Hungarian algorithm achieves this in $O(n^3)$ time.

\subsection{Attack Summary}\label{attack_summary}
The basic mechanism of the baseline attack~\cite{Xu_2017} is to iteratively match each individual's locations of the $i$th time step with those of the $(i+1)$th. Each assignment produced by the Hungarian algorithm produces the estimated locations for the next time step. The cost matrix at each time step is $C_i \in \mathbb{R}^{n \times n}$, where rows represent individuals, and columns represent locations. While there are actually $m$ locations, the columns enumerate each individual from the aggregated record $r_i$. For example, if $r_i$ has $x$ many people in location $y$, then $x$ many columns in $C_i$ shall represent location $y$. Thus, we must record which columns represent which locations for every time step. Costs are determined by heuristics based on human mobility, primarily leveraging the observation that most people have regular mobility patterns~\cite{Song_2010},~\cite{Gonzalez_2008}.

The first time step of predictions for each day can be trivially determined from the input dataset. Then, the attack is split into three stages. Recovering night-hour trajectories (\numzero{00}:\numzero{00} to \numzero{06}:\numzero{00}) is the first stage, where costs are based on physical distance; this is based on the assumption that most people are immobile during night hours. The second stage is recovering the following daytime trajectories (\numzero{06}:\numzero{00} to \num{24}:\numzero{00}), where the cost is based on a simple velocity model. At the current ($i$th) time step, given a location $p_i$ and a candidate location $\ell$, this heuristic defines the cost of $\ell$ being the next location as the distance between $\ell$ and $q$, where $q$ is the location estimated by extending the vector induced by the locations from the $(i-1)$th and $i$th time steps:
\begin{equation}
    q = p_i + (p_i - p_{i-1}).\label{eqa1}
\end{equation}
\begin{equation}
    \textit{cost}(p_i, \ell) = \textit{distance}(\ell, q).\label{eqa2}
\end{equation}
By this point, $n$ sub-trajectories of length $d$, where $d$ is the number of time steps within a single day, have been recovered for each day captured by the input dataset. For the third stage, each of these sub-trajectories must be uniquely related to each other to recover the full set of $n$ trajectories that last for all $t$ time steps. To obtain this matching, the Hungarian algorithm is again used on a cost matrix where rows represent a day's sub-trajectories and columns represent the next day's sub-trajectories. Based on the observation that people have repetitive daily movements~\cite{Gonzalez_2008}, Xu et al. use a standard formulation of information gain to measure the similarity between two sub-trajectories for cost. The three-stage attack is deterministic and runs in $O(tn^3)$ time.

\subsection{Aggregated Dataset Requirements}\label{data_requirements}
The baseline attack~\cite{Xu_2017} targets aggregated mobility datasets. Therefore, a suitable dataset must comply with certain requirements. While physical location details are still required, each location must be treated as discrete to represent an area, for example, a mobile base station to which mobile phone users are connected, as per the dataset used in~\cite{Xu_2017}. The aggregated dataset must be complete and contain records of the same set of people, i.e. the total sum within every record must equal $n$, no records are missing, and no people are unaccounted for. The interval between each time step must be regular and evenly divide \num{24} hours. The dataset records must begin between \numzero{00}:\numzero{00} and \numzero{06}:\numzero{00}.

\section{Design and Heuristic Enhancements}\label{enhancements}
We propose the following design improvements and heuristic enhancements that improve accuracy while reasonably maintaining the efficiency and determinism of the baseline and permit the online execution of the attack.

\noindent \textbf{Stage reduction.} The heuristic of the baseline attack assumes trajectories are static during night hours. However, this is not necessarily the case for every person. For example, approximately \SI{18}{\%} of taxi trips from the raw Porto Taxi dataset were recorded as beginning during night hours~\cite{Porto_Taxi}. For trajectories that do not conform to this assumption, predictions should still accurately consider movement, which, by design, this heuristic does not. Thus, we only use the static distance-based baseline heuristic to generate the location for the time step immediately after the trivial first-step prediction (for midnight) each day. The modified velocity heuristic below determines the remaining $d-2$ predictions for each day. Note that for trajectories that do conform to this static assumption, the velocity heuristic still models immobility accurately and the distribution of locations for the next time step deduced from the input dataset also accounts for this.

\noindent \textbf{Heuristic alterations.} We introduce a matrix $B \in \mathbb{N}^{m \times m}$, where each element $b_{i,j} \in B$ is the number of times location $j$ has been predicted to follow location $i$, for all time steps up to and including the final time step of the last fully-predicted day. This is akin to a bigram or transitional matrix used in algorithms to model hidden Markov processes, such as the Viterbi algorithm~\cite{Forney_1973}. Recording such information allows for future predictions to be affected by historical ones.

The original velocity heuristic is described in Section~\ref{attack_summary}. Intuitively, this heuristic is restrictive as it does not account for direction changes well, nor does it consider whether the path is common or even possible. For example, if a curved railway surrounded by isolated farmland leads to a popular airport, a linear model may estimate locations leading off the railway and into the farmland. This causes the cost calculation in~(\ref{eqa2}) to be inaccurately based on a poor estimation when an estimation based on popularity was a better choice. Such considerations are similarly helpful for cases where a linear estimation gives a completely inaccessible location, for example, offshore or mountainous. It is, therefore, natural to additionally consider the possible popularity and repetitiveness of certain locations. Inspired by hidden Markov processes, we utilise the information from $B$ by redefining the cost as:
\begin{equation}
    H_p = \{i \mid b_{p,i} = \max_{1 \le j \le m}(b_{p,j}) \land b_{p,i} > 0\}.\label{eq1}
\end{equation}
\begin{equation}
    \textit{cost}(p,\ell) = \min_{x \in H_p \cup \{q\}}\textit{distance}(\ell,x).\label{eq2}
\end{equation}
where $b_{i,j}$ represents the element in the $i$th row and $j$th column of $B$, and $q$ is defined as per~(\ref{eqa1}).

\noindent \textbf{Sub-trajectory linkage alterations.} The baseline attack links sub-trajectories of length $d$ together using information gain as the cost to match similar sub-trajectories. While human mobility patterns are expected to be regular, anomalous trajectories for certain days are still possible. Furthermore, if an incorrect day-to-day linkage is made, this severely impacts the accuracy of the entire trajectory once later linkages are made. Thus, we introduce a positive integer parameter $k$ that expresses the number of previous days to consider. With trajectory $u$ and recently predicted sub-trajectory $v$, we redefine the cost of linking them as:
\begin{equation}
\textit{cost}(u, v) = \min_{0 \le i < k}g(u_{x-i}, v).\label{eq3}
\end{equation}
where $u_{*}$ means the $u$th sub-trajectory for the $*$th day, $x$ is the last fully-predicted day, and $g$ is the information gain function, as used in the baseline attack. Note that setting $k=1$ is equivalent to the linkage mechanism of the baseline attack.

\noindent \textbf{Order of computation.} The design of the baseline attack considers these linkages as the ``third stage'', to be performed after all sub-trajectories of length $d$ are independently recovered. Our alterations require that the trajectories be recovered chronologically and linked cumulatively. After each sub-trajectory of length $d$ is predicted, they must be linked to the existing trajectories, followed by an update to the bigram matrix $B$. This modified order of computation enables the algorithm to be run online, the ramifications of which are outlined in Section~\ref{threat_model}.

With these alterations, the algorithm remains deterministic and requires minimal additional computational resources. During experimentation, it was determined that values for $k > 7$ (representing mobility pattern repetition schedules of more than one week) were of little to no benefit to the accuracy, so we reasonably assume that $k \ll m$, $k \ll n$, and that $k$ is bounded by a small constant. Then, the enhanced attack runs in $O(t(n^3 + n^2m))$ time and requires additional $O(m^2)$ space compared to the baseline attack. Note that whether $m > n$ depends on external factors of the input dataset, such as population density and spatial resolution. The results of these enhancements are presented in Section~\ref{evaluation}.

\section{Data}\label{data}
\subsection{Dataset 1: GeoLife}\label{data-geolife}
GeoLife~\cite{GeoLife}, collected by Microsoft Research Asia, is a dataset of time-stamped GPS locations from \num{182} different users in Beijing, China. The data spans from \num{2007} to \num{2012}, but most users are inactive for most of this period. It contains over \num{17000} trajectories with a total time of over \num{48000} hours. %
As mentioned in Section~\ref{data_requirements}, we require a dataset with uniform time intervals, however the raw dataset has variable time intervals. Furthermore, it is sparse, with too few users with trajectories in a common period. We apply the following manipulations to address these issues.

Most spatiotemporal points are located within a certain region of Beijing. Points outside Beijing and users with only a single trajectory are discarded. Then, only records from the top-$k$ most active months for each user are retained. To partially address the inconsistent interval of time stamps, a floor operation is applied to each time stamp, followed by removing duplicates caused by this operation. The remaining interval-related inconsistencies are resolved with interpolation (see Section~\ref{general-preprocessing}).

Additionally, some users have multiple trajectories that span only a few hours, complicating finding a common period with multiple active users. To resolve this, we shift consecutive trajectories temporally closer to each other. If these are also deemed to be reasonably close spatially, they are merged to create a longer trajectory. As one consistent trajectory is required, only the longest trajectory for each user is kept. Then, we assign the same start date-timestamp to all trajectories and adjust the trajectories to follow the new starting date. With all the trajectories aligned, all points beyond the earliest ending time-stamp are discarded, resulting in a dataset temporal coverage of approximately one week.

\subsection{Dataset 2: Porto Taxi}\label{data-portotaxi}
Porto Taxi~\cite{Porto_Taxi} is a dataset of spatiotemporal points collected from taxis in Porto, Portugal. It contains \num{1710670} trajectory records of \num{444} taxis over the span of one year, from \numzero{01}/\numzero{07}/\num{2013} to \num{30}/\numzero{06}/\num{2014}. Each trajectory represents a taxi trip and is given as a list of GPS coordinates captured at an interval of approximately \SI{15}{\second}. Some trajectories have missing points; these are immediately discarded, leaving \num{1704685} trajectories.

In addition to complying with the requirements in Section~\ref{data_requirements}, the objective is to retain a maximal dataset size while minimising our interference. To achieve this, we identify the densest period of taxi trips in the year, then retain only the trajectories of taxis that have completed an above-average number of trips within that period, thus limiting the application of interpolation described in Section~\ref{general-preprocessing}.
The monthly maximum number of taxi trips is in May, the first full week of which has the weekly maximum. After filtering as described, \num{197} trajectories remain. Then, each trajectory is sampled every \SI{10}{\minute}, resulting in \num{4321} time steps for a \num{30} day period.

\begin{figure}[b]
\vspace{-1em}
\centerline{\includegraphics[width=0.8\columnwidth]{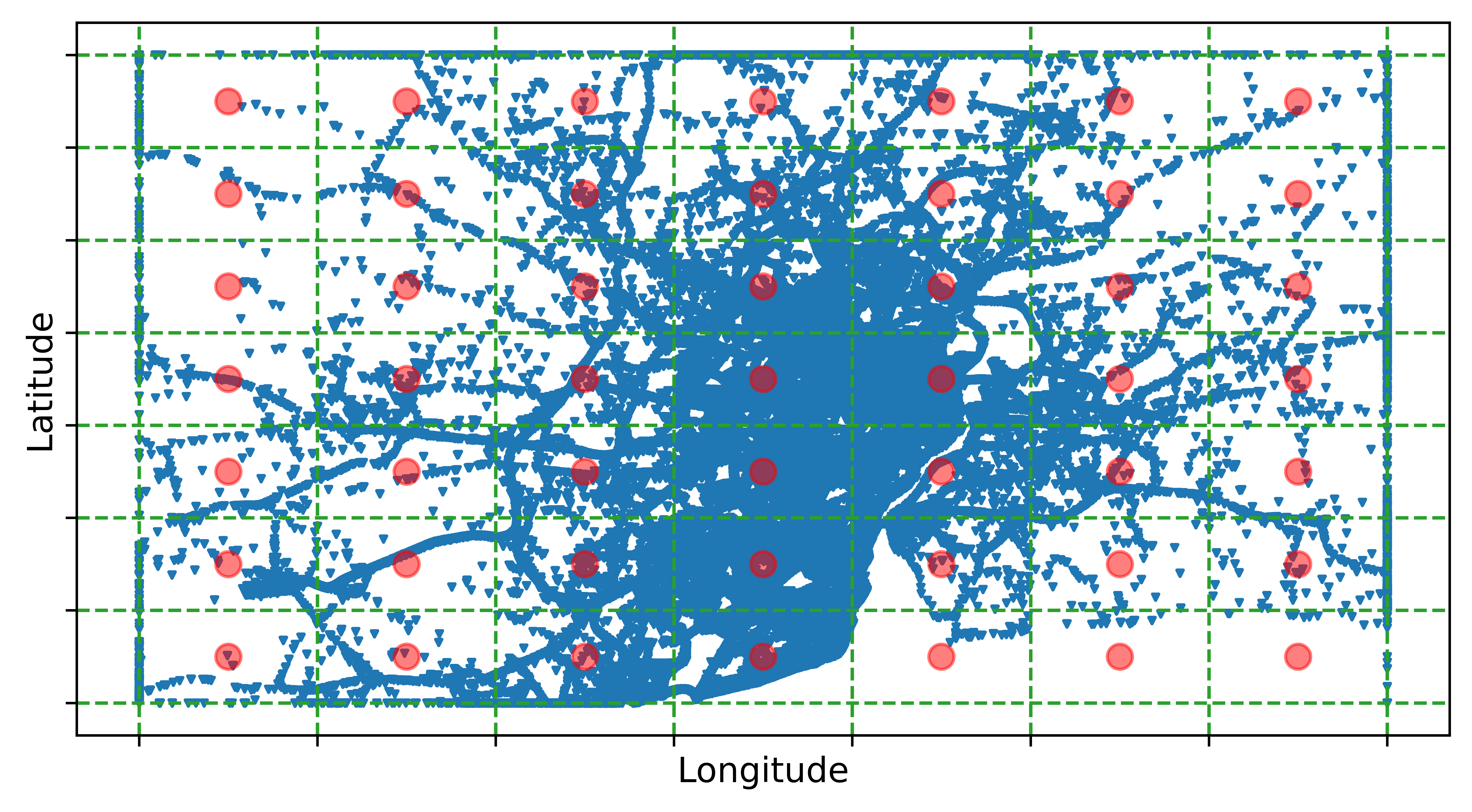}}
\vspace{-1em}
\caption{Location grid with cell centres (red) and user locations (blue) applied to the Porto Taxi dataset.}
\label{portotaxi-grid}
\end{figure}

\subsection{General Preprocessing Steps}\label{general-preprocessing}
After completing the specific preprocessing mentioned above, the following operations are required.

No location records are available for the period between two records where users are idle. This gap is addressed by filling it with interpolated records, though we must note that no interpolation technique can entirely reflect a true mobility pattern. A static interpolation method was deemed the most appropriate, where the user's last-known location is repeated for each time step until the next-known location. This was selected over other techniques, such as linear interpolation, where missing locations are filled by regularly spacing locations between the last-known and next-known locations for each intermediate time step because such a method forces trajectories to contain locations that may not exist or be accessible.

To enforce the locations as discrete areas, as shown in Fig.~\ref{portotaxi-grid}, a rectangular region is defined with the bottom-left and top-right corners at the \num{1}st and \num{99.5}th percentiles of latitude and longitude, respectively. Any spatiotemporal points outside this enforced boundary are shifted to the boundary, ensuring that the true location is represented in at least one spatial dimension. The region is then divided into square cells; points located within a cell are said to belong to that `location'. The cells were set to represent an area of \SI{1}{\kilo\metre\squared} for GeoLife, and \SI{4}{\kilo\metre\squared} for Porto Taxi, reflecting practical spacings between modern cellular network towers in urban areas~\cite{Prkic_2014}. The physical location is considered to be the centre of that cell. A user's location at any time step is taken as the ground truth location within each trajectory. The total number of users in each location at each time step is taken as aggregated data.

\subsection{Limitations}\label{data_limitations}
As outlined, extensive preprocessing was necessary to align the datasets with the attack framework's requirements (see Section~\ref{data_requirements}). We could not obtain the datasets used in the original publication due to access restrictions, and we are unaware of any other high-quality open-source datasets with similar properties that could have been used instead. Additionally, the Porto Taxi dataset specifically targets taxi drivers, whose mobility patterns differ significantly from typical mobile phone users. After preprocessing, the number of users in both datasets was considerably lower than in the original study. These constraints have inevitably impacted our results, but there were no alternatives for conducting the attack using openly available data.

\section{Implementation Details}\label{implementation}
All code is provided in Python \num{3.10} and released as open-source under the MIT licence\footref{repo}. The implementation deploys the Numpy, Pandas, Matplotlib, Scipy, Geopy, Levenshtein, and Tqdm packages.
All preprocessing code is released as annotated Jupyter Notebooks describing the process in detail.
The baseline attack~\cite{Xu_2017} was re-implemented in a Jupyter Notebook that contains extensive explanations detailing each step of the algorithm and shows intermediate results. This representation clarifies the nature of the privacy leakage and details the attack mechanisms and the features of the data that lead to the leakage.
Moreover, we provide Python classes for both the baseline and enhanced attacks that allow execution via scripts.
The modular nature of this implementation allows for (additional) datasets to be readily loaded, visualised, and evaluated with minimal additional code, facilitating further research. The repository further contains full API documentation.
The implementation of our enhanced attack allows predictions to be accessed online from another thread during execution.

\begin{table}[b]
\vspace{-1em}
\centering
\caption{Sub-dataset details}\label{tab:sub-datasets}
\begin{tabular}{p{1.8cm}p{0.8cm}p{1.2cm}p{1.5cm}p{1cm}}
\toprule
\textbf{Dataset} & \textbf{\#Users} & \textbf{\#Locations} & \textbf{\#Time Steps} & \textbf{Interval} \\
\midrule
GeoLife \num{37} & \num{37} & \num{194} & \num{5205} & \SI{2}{\minute} \\
GeoLife \num{38} & \num{38} & \num{492} & \num{10407} & \SI{1}{\minute} \\
GeoLife \num{43} & \num{43} & \num{120} & \num{10103} & \SI{1}{\minute} \\
Porto Taxi $3 \times 3$ & \num{197} & \num{9} & \num{4321} & \SI{10}{\minute} \\
Porto Taxi $7 \times 7$ & \num{197} & \num{49} & \num{4321} & \SI{10}{\minute} \\
\bottomrule
\end{tabular}
\end{table}

\section{Evaluation and Discussion}\label{evaluation}
\noindent \textbf{Metrics.} To draw comparisons, the metrics used for evaluation mirror those utilised by Xu et al.~\cite{Xu_2017}. In their study, they define \textit{accuracy} as the proportion of correctly predicted spatiotemporal points, given by:
\begin{equation}
    \textit{accuracy} = \frac{1}{n}\sum_{i=1}^{n} \frac{|A_i \cap B_i|}{t}.\label{eqr1}
\end{equation}
where $A_i$ and $B_i$ represent the $i$th predicted and corresponding true trajectories respectively. This can be described as the average of the complement of the normalised Hamming distances~\cite{Hamming_1950} between every predicted trajectory and the associated (see Mapping below) true trajectory. Additionally, they define the \textit{recovery error} as the total sum of distances between predicted and true spatiotemporal points. They also introduce the \textit{top-$k$ uniqueness}~\cite{Montjoye_2013} of a dataset as ``the percentage of recovered trajectories that can be uniquely distinguished by their most frequent $k$ locations''~\cite{Xu_2017}. This metric quantifies how easily the recovered trajectories can re-identify individuals, completing the de-anonymisation process. A natural example is $k=2$, where the two most frequent locations can be assumed to be one's home and workplace~\cite{Golle_2009}. Intuitively, high uniqueness indicates more severe levels of privacy leakage in the dataset, as individuals are more unique regarding the places they frequent and are, therefore, easier to identify.

\begin{figure}[b]
    \centering
    \vspace{-1em}
    \includegraphics[width=0.9\columnwidth]{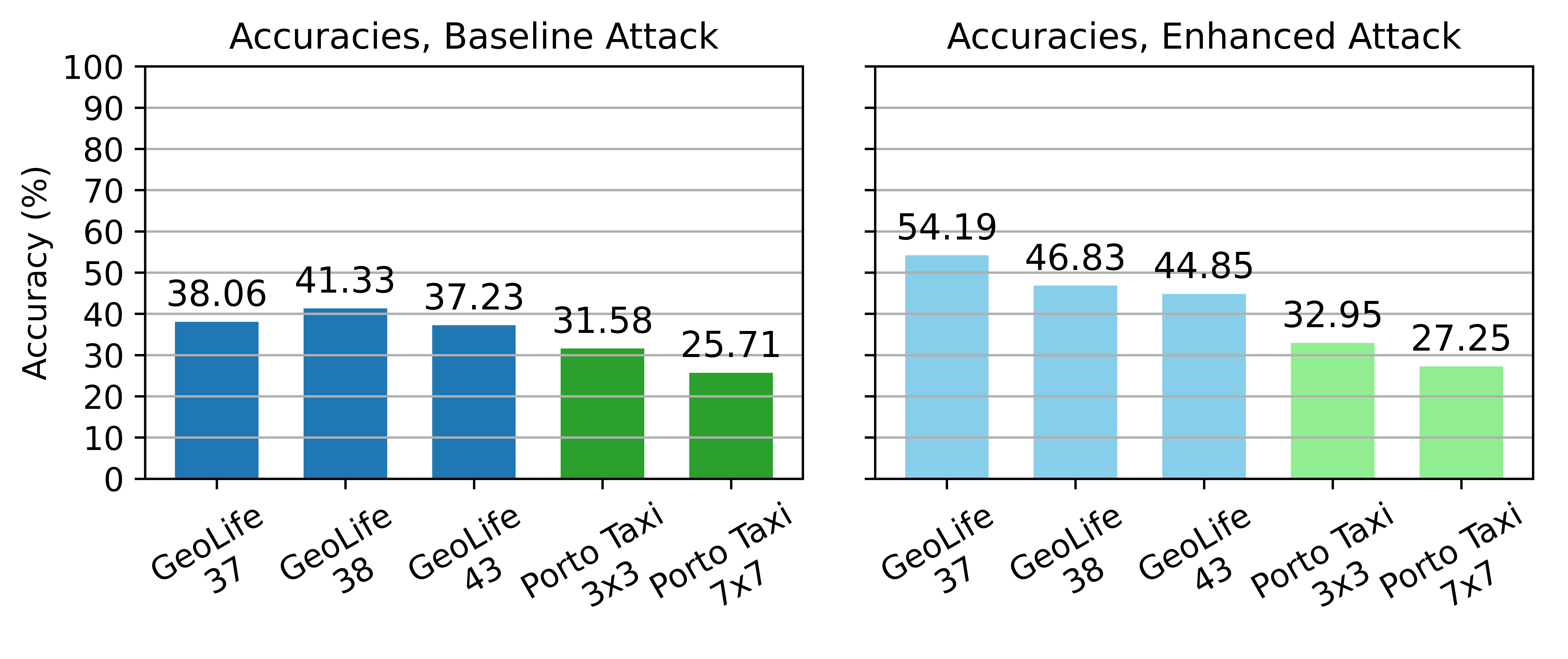}
    \vspace{-1.25em}
    \caption{Accuracies on the baseline (left) and enhanced (right) attacks.}
    \label{fig:accuracies}
\end{figure}

\noindent \textbf{Mapping.} To apply these metrics, the predicted trajectories must be associated with true trajectories by creating a one-to-one matching between the two sets. To achieve this, we create another cost matrix $C \in \mathbb{R}^{n \times n}$, where costs are defined as the recovery error between the two trajectories, and apply the Hungarian algorithm to produce this mapping. By contrast, the mapping method used in~\cite{Xu_2017} is greedy, achieved by iteratively matching each predicted trajectory with the most similar unmatched true trajectory. This potentially results in sub-optimal pairings, which negatively affects the accuracy. We opted for the Hungarian algorithm, as it ensures that the recovery error is globally minimised, maximising the accuracy. Note that the improved mapping was used for both baseline and enhanced attacks to ensure a fair comparison.

\begin{figure}[t]
    \centering
    \includegraphics[width=\columnwidth]{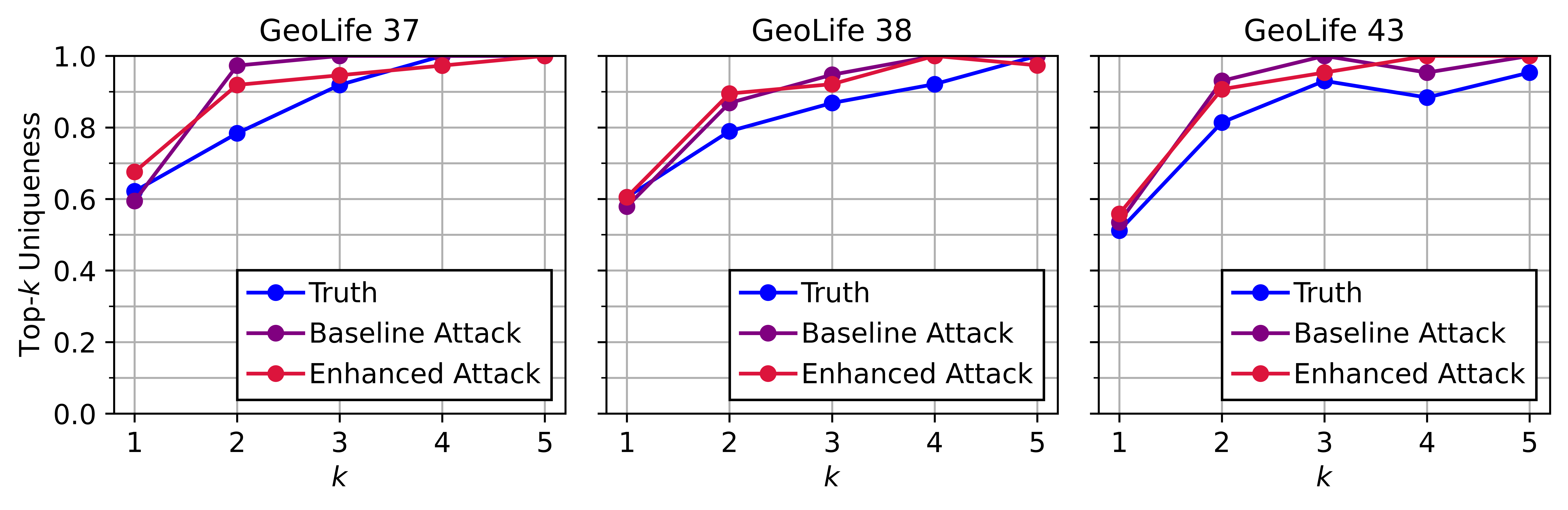}
    \includegraphics[width=0.67\columnwidth]{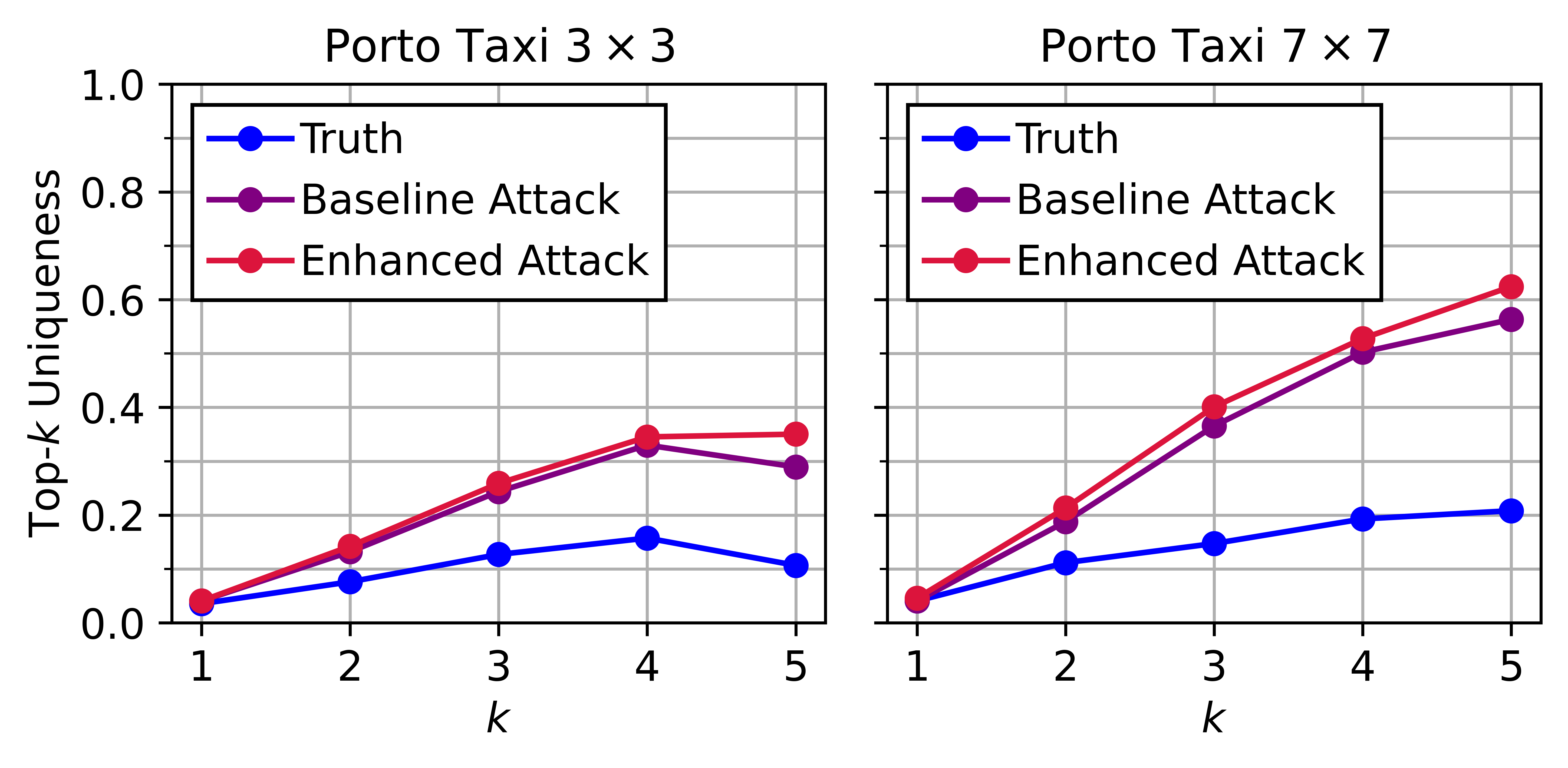}
    \vspace{-1em}
    \caption{Top-$k$ uniqueness values for $1 \le k \le 5$. The ground truth and outputs of the baseline and enhanced attacks are shown for each sub-dataset.}
    \label{fig:uniqueness}
    \vspace{-1em}
\end{figure}

\begin{figure}[b]
    \vspace{-1em}
    \centering
    \includegraphics[width=\columnwidth]{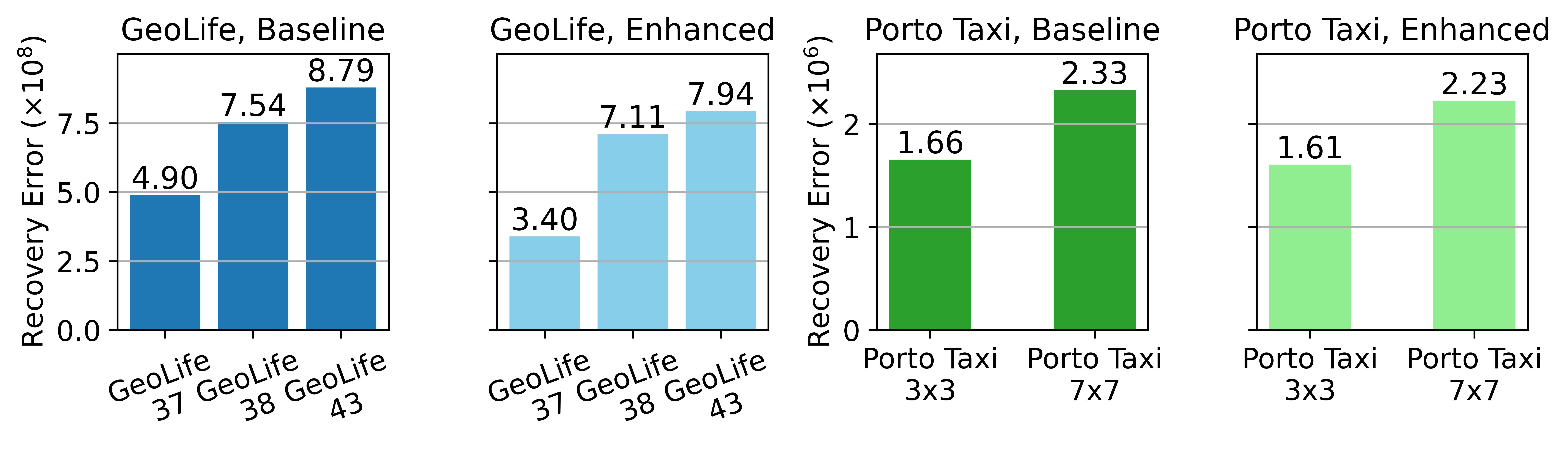}
    \vspace{-2em}
    \caption{Recovery errors on the baseline and enhanced attacks.}
    \label{fig:recovery-errors}
\end{figure}

Following the preprocessing outlined in Section~\ref{data}, we ultimately obtained three sub-datasets from GeoLife and two from Porto Taxi. We evaluated the baseline and the enhanced attack on each of these. The details of each sub-dataset are shown in Table~\ref{tab:sub-datasets}. The accuracies, recovery errors, and top-$k$ uniquenesses are shown in Figs.~\ref{fig:accuracies},~\ref{fig:recovery-errors}, and~\ref{fig:uniqueness}, respectively.

As shown in Fig.~\ref{fig:accuracies}, the attack was more successful on the GeoLife datasets. The highest accuracy achieved by the baseline attack was \SI{41}{\%} on the \num{38}-user dataset, with our enhancements increasing this to \SI{46}{\%}. The enhanced attack's highest accuracy was achieved on the \num{37}-user dataset, reaching \SI{54}{\%}, while the baseline algorithm achieved \SI{38}{\%}, highlighting the largest marginal improvement from our enhancements. More generally, the enhanced version yielded higher accuracies for every sub-dataset across both datasets, especially for those derived from GeoLife. All measurements for our enhancement were conducted with a linkage parameter of~$k=3$.

Overall, these accuracies are significantly lower than those achieved by Xu et al. on their commercial datasets, where they achieved \num{73}-\SI{91}{\%} using the baseline attack. This is partially expected given our explanations of the limitations of our datasets in Section~\ref{data_limitations}. We expect that with our much smaller datasets, the accuracy metrics suffer from granularity-related noise. We also note that GeoLife is a human mobility dataset, while Porto Taxi is a vehicular mobility dataset, and that the heuristics used in~\cite{Xu_2017} that we replicate and extend are based primarily on human mobility. This explains the lower accuracies in general from the Porto Taxi dataset and the marginally smaller improvements from the enhancements. Given that the heuristics designed by Xu et al. were based on human mobility patterns, we hypothesise that using heuristics tailored to other types of mobility should give more accurate results for those kinds of data. Fig.~\ref{fig:uniqueness} also shows that uniqueness values for GeoLife are far higher, facilitating higher accuracies. The vehicular nature of the Porto Taxi dataset may also explain its low uniqueness values. For example, popular points of interest (such as an airport) and routes (such as arterial highways) reduce the uniqueness of the data, resulting in less accurate trajectories, as shown by the results. However, the results of the GeoLife dataset suggest that our enhancements can recover trajectories more accurately than the original attack.

\begin{figure}[t]
    \centering
    \includegraphics[width=0.7\columnwidth]{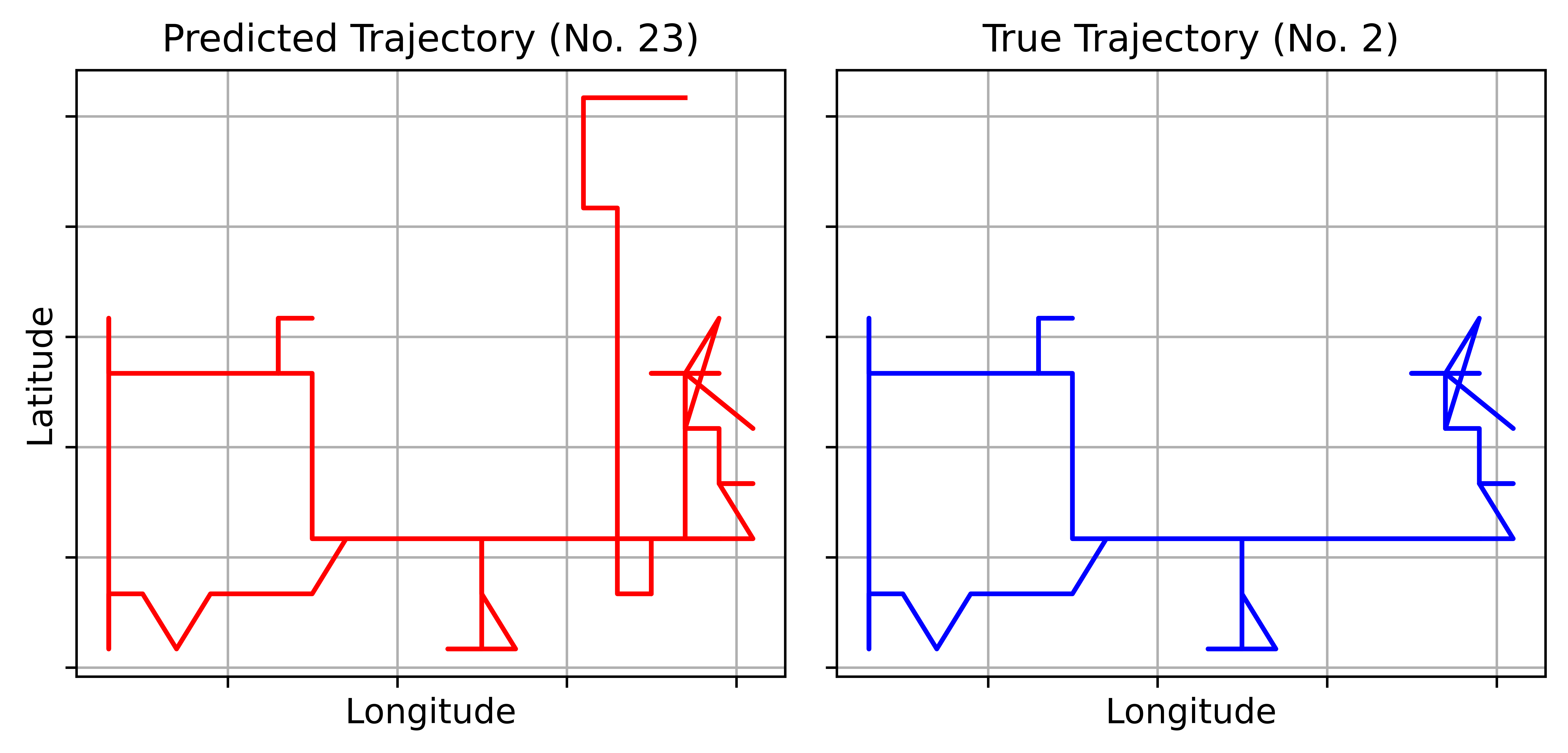}
    \hspace*{0.125cm}\includegraphics[width=0.7\columnwidth]{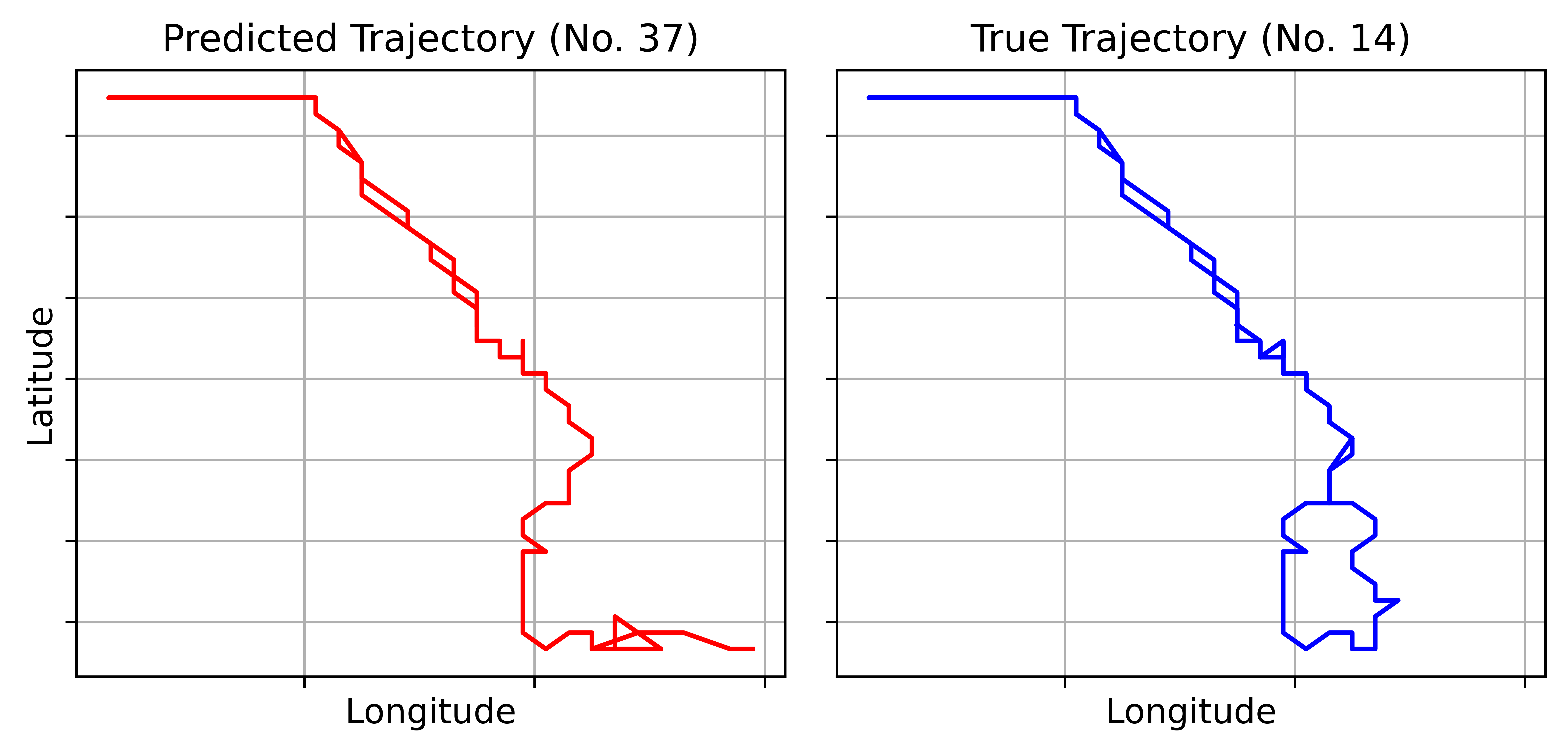}
    \vspace{-0.75em}
    \caption{Two examples of predicted trajectories and their matched true trajectories over a single day.}
    \label{fig:prediction-1}
    \vspace{-1em}
\end{figure}

We also observe the trajectories predicted by the attack compared to their associated true trajectories in Fig.~\ref{fig:prediction-1}. In both examples, the trajectories are mostly recovered, apart from some outlying patterns. In situations where the locations are not exactly correct, we also observe that the heuristics described in~\cite{Xu_2017} must somewhat accurately capture patterns in human mobility. Hence, we conclude that there is some privacy leakage from the processed aggregated datasets.

A downside of using the Hamming distance in an accuracy measure is the possibility of a trajectory matching attaining poor accuracy due to minor errors in one of the spatiotemporal dimensions. An example of such a situation occurs when a trajectory contains an almost perfect sub-sequence of locations but is incorrectly shifted one time step. In terms of edit distance operations, the Hamming distance only permits substitutions~\cite{Navarro_2001}. To additionally consider insertions and deletions, we evaluated the Levenshtein distance~\cite{Levenshtein_1966} and used it as an alternative measure of accuracy as follows:
\begin{equation}
    \textit{Levenshtein accuracy} = \frac{1}{n}\sum_{i=1}^{n}1 - \frac{L(A_i, B_i)}{t}.\label{eqlev}
\end{equation}
where $L$ represents the Levenshtein distance function.

\begin{figure}[t]
    \centering
    \includegraphics[width=0.9\columnwidth]{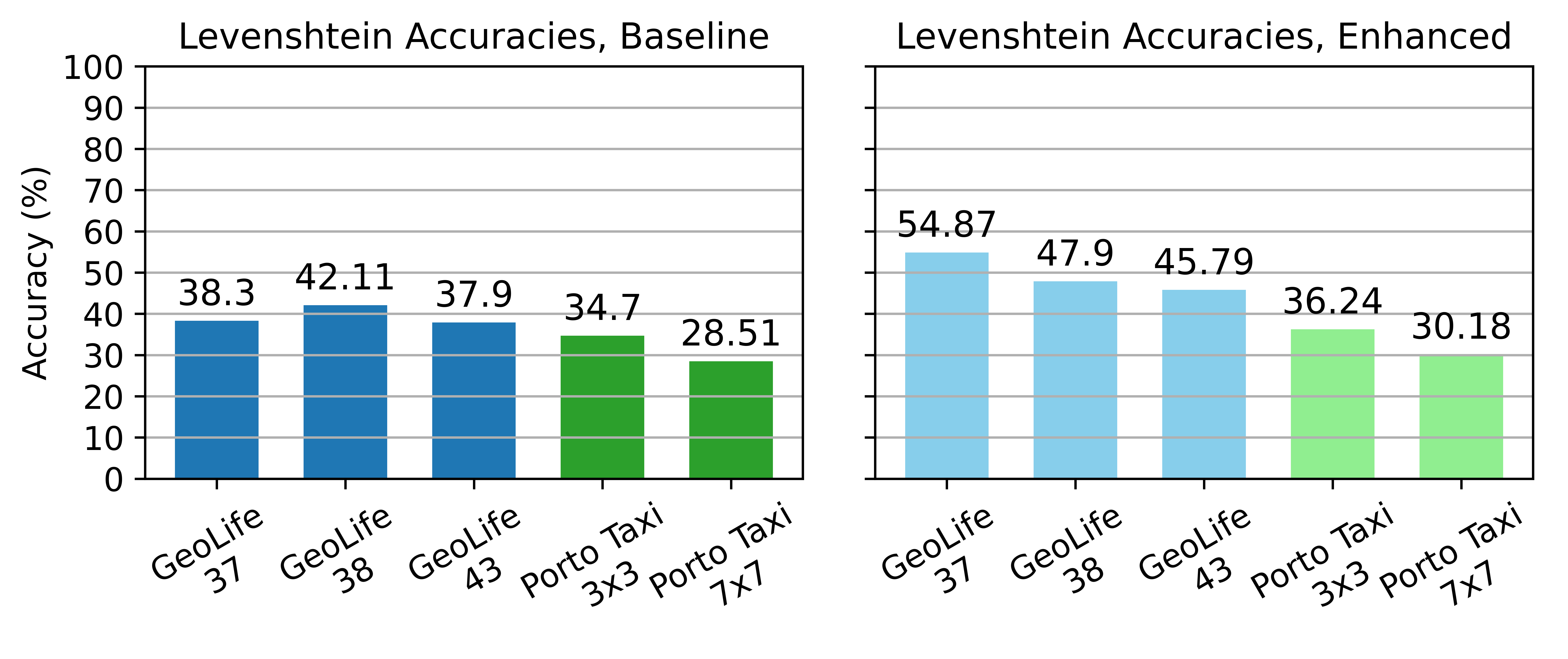}
    \vspace{-1.25em}
    \caption{Levenshtein accuracies on baseline (left) and enhanced (right) attacks.}
    \label{fig:levenshtein}
    \vspace{-1em}
\end{figure}

Fig.~\ref{fig:levenshtein} shows the Levenshtein metrics on each sub-dataset with both attack versions. The results each evaluate slightly higher than the accuracies shown in Fig.~\ref{fig:accuracies}, but generally follow the same profile, further confirming the reliability of these metrics on these datasets.

Although these results show that the recovery of trajectories is possible with the baseline attack, they suggest that initial claims about the accuracy and, therefore, the severity of privacy leakage may have been overly ambitious. Despite GeoLife being a human mobility dataset and preprocessed in such a way as to mimic the commercial mobile operator dataset used in~\cite{Xu_2017} as closely as possible, there is a significant disparity between the results, even when including our enhancements. This demonstrates a lack of ability for the attack to generalise to other datasets and highlights that the specific datasets used in~\cite{Xu_2017} are exceptionally well-suited for the task. Based on our evaluations, we claim that accuracies of up to \SI{91}{\%} somewhat misrepresent the capabilities of such an attack in real-world scenarios. The baseline attack's outstanding performance in \cite{Xu_2017} leans on strong assumptions for the considered dataset and the contained users. The inaccessible datasets used by the baseline contained many trajectories with fine-granular sampling over long periods of time for a fixed set of users. Moreover, the users seemed to comply with the strong assumptions about human mobility made by the authors, such as very limited movement during night hours. While reporting leakage in a worst-case setting is important, such assumptions do not transfer to real-world datasets that commonly suffer from less regular and uniform samples, such as the considered GeoLife and Porto Taxi datasets. Thus, the remarkably high accuracies exceeding \SI{90}{\%} in the baseline work~\cite{Xu_2017} might significantly overestimate the success of the attack in practical scenarios. Nevertheless, the results confirm that privacy leakage is a real concern, and the risk should not be underestimated.

\section{Future Work}\label{future}
Throughout previous sections, we established how heuristics significantly influence the performance of this attack. Naturally, it is interesting to investigate whether using heuristics tailored to other forms of mobility results in similar privacy leakages for other types of mobility data. Further experimentation on more human mobility datasets (as they become available) will also clarify the disparity between the results achieved on our public open-source datasets and the private commercial ones evaluated by Xu et al.~\cite{Xu_2017}.

The heuristic methods used by us and in~\cite{Xu_2017} are limited in their ability to fully encapsulate the dynamic nature of human movements because they are based on researcher-defined assumptions that potentially overgeneralise the more complex patterns evident in human mobility~\cite{Gonzalez_2008}. Prior research in location trajectory privacy suggests that deep-learning methods may be able to address this. For example, Wang et al.~\cite{Wang_2019} explore the usage of LSTMs and Seq2Seq approaches for the trajectory prediction task, and multiple authors~\cite{Gao_2017, marc2020} leverage Recurrent Neural Networks for trajectory user linking. Their results illustrate the potential of deep learning to improve accuracy and robustness over static heuristic methods.

\section{Conclusion}\label{conclusion}
To evaluate the privacy leakage of aggregated mobility datasets, we successfully reimplemented the trajectory recovery attack proposed by Xu et al.~\cite{Xu_2017}. The original study evaluated the attack on inaccessible commercial datasets, rendering the results irreplicable and the subsequent claims unverifiable. To increase transparency in this area of research, we initially conducted the same attack with our reimplementation, using public open-source datasets, namely GeoLife~\cite{GeoLife} and Porto Taxi~\cite{Porto_Taxi}. To further facilitate future research, we designed improvements to the baseline that yielded substantially higher accuracies (by up to \SI{16}{\%}), requiring minimal additional computation, for use as an improved baseline. Our improvements also permit an online version of the attack, making the attack significantly more accessible to larger datasets previously considered unprocessable. We released all code as open-source to ensure our findings are reproducible. Our results, attaining accuracies of up to \SI{54}{\%} on the GeoLife dataset and \SI{32}{\%} on the Porto Taxi dataset, show that the reconstruction of individual trajectories from anonymised aggregated data represents a practical risk. The results confirm the privacy concerns raised by Xu et al. but also suggest that the originally reported results are over-exaggerated and depend on strong assumptions about the considered dataset and users. Nevertheless, this work emphasises the need for enhanced privacy protection measures when publishing aggregated mobility data.

\ifreview
\else
\section*{Acknowledgment}
    The authors would like to thank UNSW, the Commonwealth of Australia,
    and the Cybersecurity Cooperative Research
    Centre Limited for their support of this work.
\fi

\bibliographystyle{ieeetr}
\bibliography{references}
\end{document}